\documentclass[aps,prl,reprint,showpacs,superscriptaddress,longbibliography]{revtex4-2}
\usepackage{mathtools,amssymb,graphicx,units}

\usepackage[plainpages=false,pdfpagelabels,colorlinks=true,linkcolor=red,urlcolor=blue,citecolor=blue,pdftitle={Title},pdfauthor={},pdfdisplaydoctitle=true,pdfduplex=DuplexFlipLongEdge]{hyperref}
\usepackage{verbatim}
\usepackage[english]{babel}
\usepackage{color}
\definecolor{darkred}{rgb}{0.80,0,0}
\definecolor{blood}{rgb}{0.50,0,0}
\definecolor{brightred}{rgb}{1,0,0}
\definecolor{orange}{rgb}{1,0.3,0}
\definecolor{bluegreen}{rgb}{0,0.5,0.5}
\definecolor{lightblue}{rgb}{0,0.5,0.8}
\definecolor{darkgreen}{rgb}{0,0.5,0}
\definecolor{green}{rgb}{0,0.70,0}
\definecolor{darkblue}{rgb}{0,0,0.80}
\definecolor{magenta}{rgb}{1,0,1}
\definecolor{softmagenta}{rgb}{0.85,0.1,0.6}
\definecolor{mauve}{rgb}{0.6,0.1,1}
\definecolor{white}{rgb}{1,1,1}
\definecolor{black}{rgb}{0,0,0}

\usepackage{ulem}

\newcommand{\beginsupplement}{%
        \setcounter{table}{0}
        \renewcommand{\thetable}{S\arabic{table}}
        \setcounter{figure}{0}
        \renewcommand{\thefigure}{S\arabic{figure}}
        \setcounter{equation}{0}
        \renewcommand{\theequation}{S\arabic{equation}}
}

\begin{document}
\title{Hamming Distance and the onset of quantum criticality}
\author{Tian-Cheng Yi}
\affiliation{Beijing Computational Science Research Center, Beijing 100193, China}
\author{Richard T. Scalettar}
\email{scalettar@physics.ucdavis.edu}
\affiliation{Department of Physics, University of California,
Davis, CA 95616, USA}
\author{Rubem Mondaini}
\email{rmondaini@csrc.ac.cn}
\affiliation{Beijing Computational Science Research Center, Beijing 100193, China}

\begin{abstract}
Simulating models for quantum correlated matter unveils the inherent limitations of deterministic classical computations. In particular, in the case of quantum Monte Carlo methods, this is manifested by the emergence of negative weight configurations in the sampling, that is, the sign problem (SP). There have been several recent calculations which exploit the SP to locate underlying critical behavior. Here, utilizing a metric that quantifies phase-space ergodicity in such sampling, the Hamming distance, we suggest a significant advance on these ideas to extract the location of quantum critical points in various fermionic models, in spite of the presence of a severe SP. Combined with other methods, exact diagonalization in our case, it elucidates both the nature of the different phases as well as their location, as we demonstrate explicitly for the honeycomb and triangular Hubbard models, in both their U(1) and SU(2) forms. Our approach charts a path to circumvent inherent limitations imposed by the SP, allowing the exploration of the phase diagram of a variety of fermionic quantum models hitherto considered to be impractical via quantum Monte Carlo simulations.
\end{abstract}

\maketitle

\paragraph{Introduction.---}
Extracting unbiased properties of quantum many-body systems exposes the challenge that numerical simulations in classical computers face in exploring quantum matter. Roughly put, one is trapped in a tale of two exponentials.  On one side, a constraint  arises due to the `exponential wall' associated with the growing dimension of the Hilbert space with the system size.  On the other side, in avoiding retrieving exact quantum many-body wavefunctions and settling instead for a statistical estimation of physical quantities, one ends up facing the sign problem~\cite{Loh1990}, which also leads to an exponential scaling of simulation times. Apart from some special (albeit important) limits such as the half-filled fermion Hubbard model~\cite{Hirsch1985}, 
the latter `wall' appears to be 
a generic (unavoidable) characteristic of quantum Monte Carlo (QMC) methods for fermionic and frustrated bosonic systems, and is conjectured to be NP-hard~\cite{Troyer2005}.

While recent studies have in fact suggested that the average sign of weights in the latter already pinpoints the regimes of strong quantum fluctuations~\cite{Wessel2017, Mondaini2021, sign-prob3, sign-prob4}, here we focus on other statistical properties that also aid in locating quantum phase transitions, irrespective of the presence of a sign problem. In particular, we investigate a specific class of QMC methods for $d$-dimensional fermionic systems, referred to as auxiliary field QMC~\cite{Blankenbecler1981, Hirsch1985, Loh1992}, which provides a framework to stochastically average observables by sampling a fictitious field in $d+1$-dimensions, introduced in a path integral formulation of the partition function. Dubbed the Hubbard-Stratonovich (HS) field~\cite{Stratonovich57,Hubbard59,Hirsch1983,Sorella91}, $s_{i,\tau}$ carries both space and imaginary time labels and decouples the interactions, allowing an exact integration of the fermionic degrees of freedom (see Methods). The statistical properties of $s_{i,\tau}$ are the central object of our analysis.

\begin{figure*}[th!]
\centering
\includegraphics[width=0.99\textwidth]{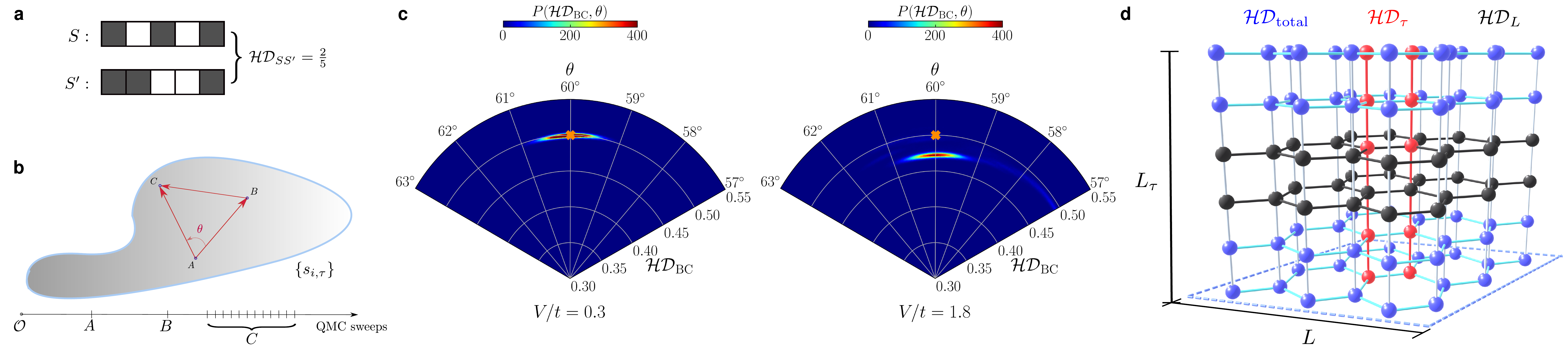}
\caption{\textbf{Hamming distance and phase space exploration.}
\textbf{a}, Representation of the Hamming distance between a pair of binary strings with length 5. \textbf{b}, Cartoon illustrating the phase space of auxiliary field configurations, accompanied by the protocol used to extract ${\cal HD}$ and a \textit{similarity} angle $\theta$ in respect to fixed points $A$ and $B$, selected after two consecutive warm-up processes. \textbf{c}, The corresponding polar probability density of BC Hamming distances for interactions far below (left) and far above (right) the known critical point of the U(1) Hubbard model on the honeycomb lattice, $V_c/t\simeq 1.35$~\cite{LeiWang2014,Li2015}. Cross markers depict the uncorrelated point, $\theta = 60^{\circ}$ and ${\cal HD}_{BC} = 0.5$. \textbf{d}, Schematic representation of the 3d lattice, which arises after the introduction of an auxiliary field $\{S\}$ to decouple the interactions; here on a honeycomb spatial lattice with linear size $L$ and $L_\tau$ imaginary-time slices. In the case of local interactions, the discrete field $\{s_{i,\tau} \}$ lives on the sites (spheres); for non-local ones, it resides on the bonds connecting orbitals $i$ and $j$, $\{s_{ij,\tau}\}$. Different colored spheres and bonds help identify the three types of Hamming distance we compute: Total (${\cal HD}_{\rm total}$), where all fields $\{s_{i\tau}\}$ are considered, ${\cal HD}_L$ where we restrict auxiliary-field `strings' to imaginary-time slices $\tau=\beta/2$, and ${\cal HD}_\tau$ where the HS field on a single site (or unit cell) across the different $L_\tau$'s is monitored. Color code schematically identify those. In \textbf{c,} $L=12$ and $L_\tau=240$.}

\label{fig:Fig1}
\end{figure*}

Recent approaches classified as `machine learning' methods, including convolutional neural networks for pattern recognition~\cite{Broecker2017,Chng17,Chng18} or clustering methods~\cite{Tiago2021a,Tiago2021b,Natanael2021}, have been applied with the aim of estimating the location of quantum critical points of many-body models using either the HS field or metrics related to it (such as matrix elements of the fermion Green's function) as an input. The fundamental observation of this paper is that the sampled Hamming distance, a simple quantity useful for establishing a separation of two points in the multidimensional phase space of auxiliary field configurations (which we take as discrete, see Fig.~\ref{fig:Fig1}\textbf{a}), already contains information regarding the onset of an ordered phase. 

The Hamming distance is generally defined as a metric for comparing two equal-length data strings, quantifying an element-wise deviation between them. In the case of \textit{binary} strings, it can be written in terms of their inner product ${\cal HD}_{x,x^\prime} = [1 - \langle x|x^\prime\rangle/\ell]/2$, where $\ell$ is the length of the string and $\langle \cdot | \cdot\rangle$ the standard vector dot product. Identical and opposite (i.e.~parity reversed) strings $x,x^\prime$ result in ${\cal HD}_{x,x^\prime} = 0$ and ${\cal HD}_{x,x^\prime}=1$, respectively, whereas completely uncorrelated strings are on average ${\cal HD}_{x,x^\prime} = 1/2$  apart. In the case of the HS field for local interactions, $\ell = N_s L_\tau$, where $N_s$ is the number of orbitals in the real-space lattice, and $L_\tau$ gives the number of imaginary-time slices in the path integral discretization of inverse temperature $\beta = \Delta \tau L_\tau$.

Non-local interactions, especially nearest-neighbor ones, lead to an HS field that resides on the bonds connecting different orbitals~\cite{Buendia1986}, whose total number we denote by $N_b$. Thus the \textit{volume} of the phase space composed by binary strings in these two cases is given by either $2^{N_s L_\tau}$ or $2^{N_b L_\tau}$. Typical importance samplings span a very small region of this vast phase space, but as we shall see, physical aspects of the model under consideration steer the sampling to correlated configurations within ordered phases, allowing one to quantitatively infer their onset.

This becomes apparent by recalling that the HS field, and in particular correlations between its constituents, serves as a proxy for correlations in real space. As demonstrated by Hirsch for the Hubbard model~\cite{Hirsch1983,Hirsch1986}, the inter-orbital fermionic spin correlations are directly proportional to the inter-spin correlations of the auxiliary bosonic field, $\langle s_{i,\tau} s_{j,0} \rangle$, with the proportionality constant $\alpha = [1 - \exp(-\Delta\tau U)]^{-1}$, and $U$ the strength of the electron-electron interaction. 
On an extreme case, when heading towards the atomic limit ($U\to\infty$), for example, the fermionic spin correlations have a one-to-one mapping to the correlations among the `spins' of the HS field (i.e.~$\alpha =1$), provided that the convergence to the continuous of the path integral discretization is slower than the one-site limit is approached \footnote{We note that the observation that $\alpha \to1$ when $\Delta\tau U \to \infty$ is still compatible to the \textit{single} approximation employed in the QMC, namely, the Trotter approximation, which gives rise to a controllable error $\propto U (\Delta\tau)^2$.}. In the case of one fermion per lattice site, the effective Hamiltonian leads to a spin pattern mimicking a Néel state in this regime. As a consequence, field configurations that follow this spin texture have a much larger weight in the sampling, driving it to a vanishingly small region of the phase space~\cite{Scalettar1991}.

To understand how this reasoning translates to regimes far from the classical one,
we investigate two models, the U(1) and SU(2) Hubbard models, in two different geometries, honeycomb and triangular lattices. The former serves as a benchmark, due to the absence of the sign problem either in the single-particle (spinful fermions) or in the Majorana representation (spinless fermions). As a result, the quantum critical points separating unordered and ordered phases are well established~\cite{Paiva05,Meng10,Sorella2012,Assaad2013,Toldin2015,Otsuka2016,LeiWang2014,Li2015}.
We then build on that benchmark and show that an investigation of the model on a triangular lattice allows us to predict the location of quantum critical points which are mostly under debate. Importantly, these are the cases where the sign problem is most severe, and hence known results come from methods that try to conquer the first exponential, either via bounding the entanglement as in matrix product-states methods in quasi-one dimensional geometries~\cite{Shirakawa2017,Szasz2020,Wietek2021,Chen2021} or via exact diagonalization (ED) in small lattices~\cite{Hotta2006,Koretsune2007,Miyazaki2009}, including its cluster derivatives~\cite{Sahebsara2008,Laubach2015}.

\paragraph{Models.---}
We investigate the spinful,
\begin{eqnarray}
\hat H = -t\sum_{\langle i j\rangle\sigma} \hat c_{i\sigma}^\dagger \hat c^{ }_{j\sigma} 
- \mu\sum_{i\sigma} \hat n^{ }_{i\sigma}
+ U\sum_i \hat n^{ }_{i\uparrow}\hat n^{ }_{i\downarrow}
\label{eq:H_spinful}
\end{eqnarray}
and the spinless,
\begin{eqnarray}
\hat H = -t\sum_{\langle i j\rangle} \hat c_{i}^\dagger \hat c^{ }_{j} - \mu\sum_i \hat n^{ }_{i}
+ V\sum_{\langle i j \rangle} \hat n^{ }_{i}\hat n^{ }_{j},
\label{eq:H_spinless}
\end{eqnarray}
Hubbard Hamiltonians, where $\hat c^{ }_{i\sigma}$ ($\hat c^{ }_{i})$ is the pseudospin-$\sigma$ (spinless) fermion annihilation operator on site $i$ and $\hat n_{i\sigma} = \hat c^{\dagger}_{i\sigma} \hat c^{ }_{i\sigma}$ ($\hat n_i=\hat c_i^\dagger c^{ }_i$) is the corresponding number density operator. Nearest-neighbor hoppings, chemical potential and repulsive interactions are given by $t$, $\mu$ and $U$ ($V$), respectively. The honeycomb and triangular geometries have a total number of sites $N_s=2L^2$, and $N_s=L^2$; imaginary-time discretization is set at $\Delta \tau = 0.1$. 

\begin{figure}[th!]
\centering
\includegraphics[width=0.99\columnwidth]{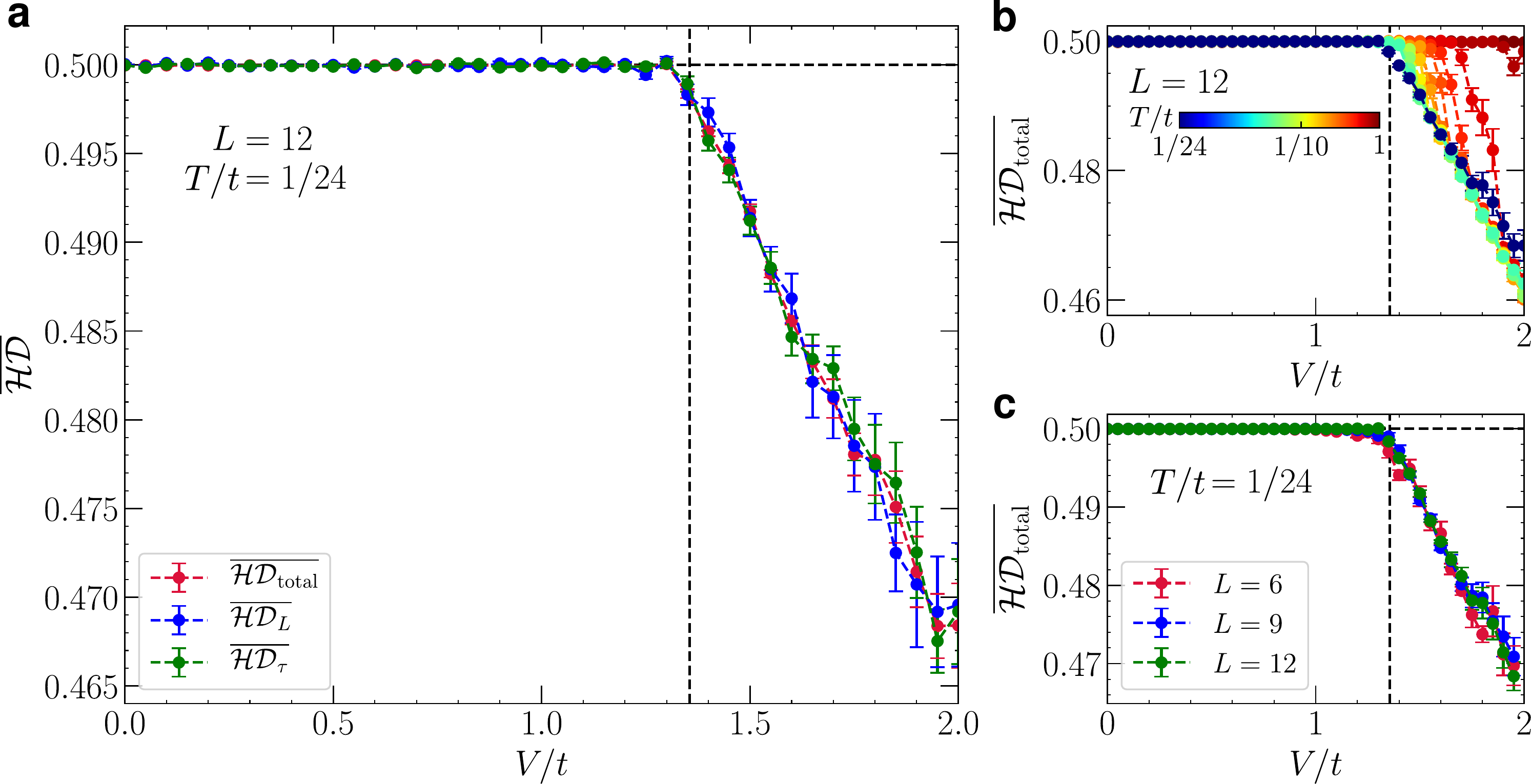}
\caption{\textbf{Hamming distance for the spinless honeycomb Hubbard model}. \textbf{a}, The three types of Hamming distance (see text), ${\cal HD}_{\rm total}$, ${\cal HD}_L$ and ${\cal HD}_\tau$ {\it{v.s.}} $V/t$, as marked. 
Vertical dashed line displays the QCP obtained in Ref.~\cite{Li2015}, $V_c/t=1.355\pm 0.001$. \textbf{b,} Temperature dependence of ${\cal HD}_{\rm total}$ on the interactions with $L=12$. \textbf{c}, Finite-size comparison of the total Hamming distance {\it{v.s.}} $V$ at $T/t=1/24$. For this case with a discrete symmetry breaking, the thermal transition can also be seen at values of the interactions that ${\cal HD}$ departs from 1/2 at finite temperatures, see Supplementary information (SI). Error bars depict the standard error of the mean (s.e.m.) over 48 independent Markov chains.
}
\label{fig:Fig2}
\end{figure}

The models on the (bipartite) honeycomb lattice are investigated at $\mu=U/2$ and $3V$, which yields `half-filling' in the SU(2) and U(1) versions of the Hamiltonian, respectively. For the triangular lattice, on the other hand, the chemical potential is systematically tuned in the  $[T\equiv1/\beta,U(V),N_s]$ set of parameters to yield one fermion per site in the spinful formulation, and one fermion per elemental triangle in its spinless version. The latter is chosen such as to render a sharp quantum phase transition to a 1/3-filled CDW state that emerges as the interactions $V$ are increased.

\paragraph{The U(1) honeycomb Hubbard model.---}
The computation of the Hamming distance between configurations of the phase space is equivalent to a projection $2^{N_b L_\tau} \to 1$ on the number of degrees of freedom. This scaling-down process is prone to miss significant features of the sampled fields, and eventually not fully characterize what is driving the sampling to become correlated. An improvement in this approach is to investigate projections to \textit{two} degrees of freedom instead. For that we take two points in the generated Markov chain, each after a significant warmup in the QMC sweeps. Thus proceeding with the usual importance sampling, while storing the distances between such points in the phase space, as illustrated in Fig.~\ref{fig:Fig1}\textbf{b}, one can define a similarity metric between configurations, i.e., an angle $\theta$ in phase space encompassed by the Hamming distance between two points in respect to a third one. Uncorrelated configurations form equilateral triangles in $\{s_{ij,\tau}\}$ (i.e., $\theta = 60^{\circ}$), and deviations from this signal a certain degree of correlations in the sampling.

Figure \ref{fig:Fig1}\textbf{c} tests this for the case of the U(1) honeycomb Hubbard model, at interactions far above and far below the known critical point $V_c/t \simeq 1.35$~\cite{LeiWang2014,Li2015}, separating a Dirac semi-metal from a charge density-wave (CDW) Mott insulator. While the majority of the angles still denote uncorrelation at either side of the transition, the typical Hamming distance significantly departs from 1/2 for $V>V_c$. For that reason, we hereafter focus primarily on $\overline{\cal HD} \equiv \overline{\cal HD}_{BC}$, the average Hamming distance, aiming in observing a signature of the known QCP location for this model. In addition, we also compute the average Hamming distance selecting fields within a fixed real space unit cell across $L_\tau$ (${\cal HD}_\tau$) or within fixed imaginary-time `layers' over real space (${\cal HD}_L$) --  see Fig.~\ref{fig:Fig1}\textbf{d} for a schematic representation. The goal is to understand if the non-ergodic behavior in the sampling has a preferential `freezing' dimension.


Figure~\ref{fig:Fig2}\textbf{a} exhibits these different quantities for a lattice with linear size $L=12$ at $T/t=1/24$. Remarkably, a sharp departure from ${\cal HD}=1/2$ is obtained around $V_c/t$, a feature largely system size independent (Fig.~\ref{fig:Fig2}\textbf{c}) when approaching the $T\to0$ limit (Fig.~\ref{fig:Fig2}\textbf{b}). Furthermore, except for minor statistical fluctuations, no significant deviations are found between the different types of averaged Hamming distances in this model, as the onset of non-ergodicity simultaneously occur in all three.

\begin{figure}[t!]
\centering
\includegraphics[width=0.99\columnwidth]{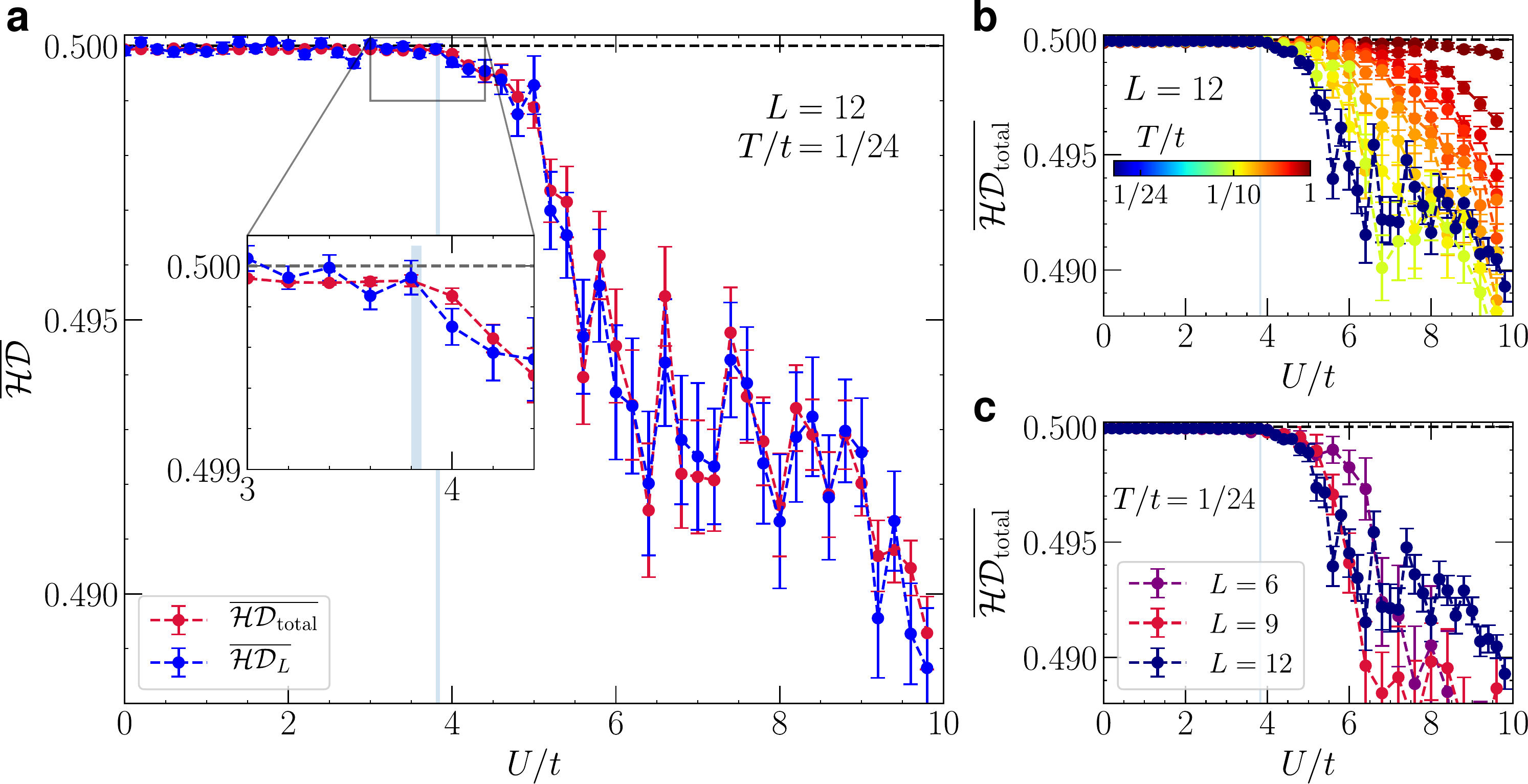}
\caption{\textbf{Hamming distance for the spinful honeycomb Hubbard model}. \textbf{a}, Total and the $\tau=\beta/2$ Hamming distances when the local interactions $U$ are swept for a honeycomb lattice with $N_s = 288$ sites 
at $T/t =1/24$. Inset displays a zoom-in on the region close to the know QCP for this model. \textbf{b} and \textbf{c} give the temperature and system-size dependence, respectively, with increasing $U$ for ${\cal HD}_{\rm total}$. Vertical shaded region depicts a confidence region of the QCP based on recent results in the literature~\cite{Sorella2012, Assaad2013, Toldin2015, Otsuka2016}. Error bars display the s.e.m. for 48 independent realizations.}
\label{fig:Fig3}
\end{figure}

\paragraph{The SU(2) honeycomb Hubbard model.---} In analogy to the spinless version, its spinful generalization features the onset of a Mott insulator at sufficiently large (local) interactions, supplanting a Dirac semi-metal phase. The insulating phase, however, exhibits a  spin-density wave (SDW), i.e., antiferromagnetic order that is triggered at $U_c/t \simeq 3.8$~\cite{Sorella2012,Assaad2013,Toldin2015,Otsuka2016}. Figure \ref{fig:Fig3} displays the equivalent of Fig.~\ref{fig:Fig2} for this Hamiltonian. Apart from larger fluctuations (even more pronounced for ${\cal HD}_\tau$, see SI~\cite{SI}), and a less marked deviation from the uncorrelated sampling regime, the Hamming distance similarly tracks the onset of the ordered phase.

Strong fluctuations on this spinful case can be interpreted by means of the larger cardinality of local degrees of freedom (four instead of two for the spinless Hamiltonian), and that the same bosonic field $\{s_{i,\tau}\}$ couples to \textit{both} fermionic flavors. Moreover, as the Mermin-Wagner theorem states that the formation of long-range magnetic order on a system with continuous symmetry at $T\neq 0$ is precluded for $d\leq2$~\cite{MerminWagner1966,Hohenberg1967}, any finite-temperature departure of ${\cal HD} = 1/2$ is understood in terms of the minimum temperature at which the quickly-decaying correlations reach typical correlation lengths comparable to the system size. This is not the case for its U(1) counterpart with a discrete symmetry, in which a finite-$T$ transition signified by the loci where ${\cal HD} < 1/2$ quantitatively matches known results for this model (See SI~\cite{SI}).

\begin{figure}[t!]
\centering
\includegraphics[width=0.99\columnwidth]{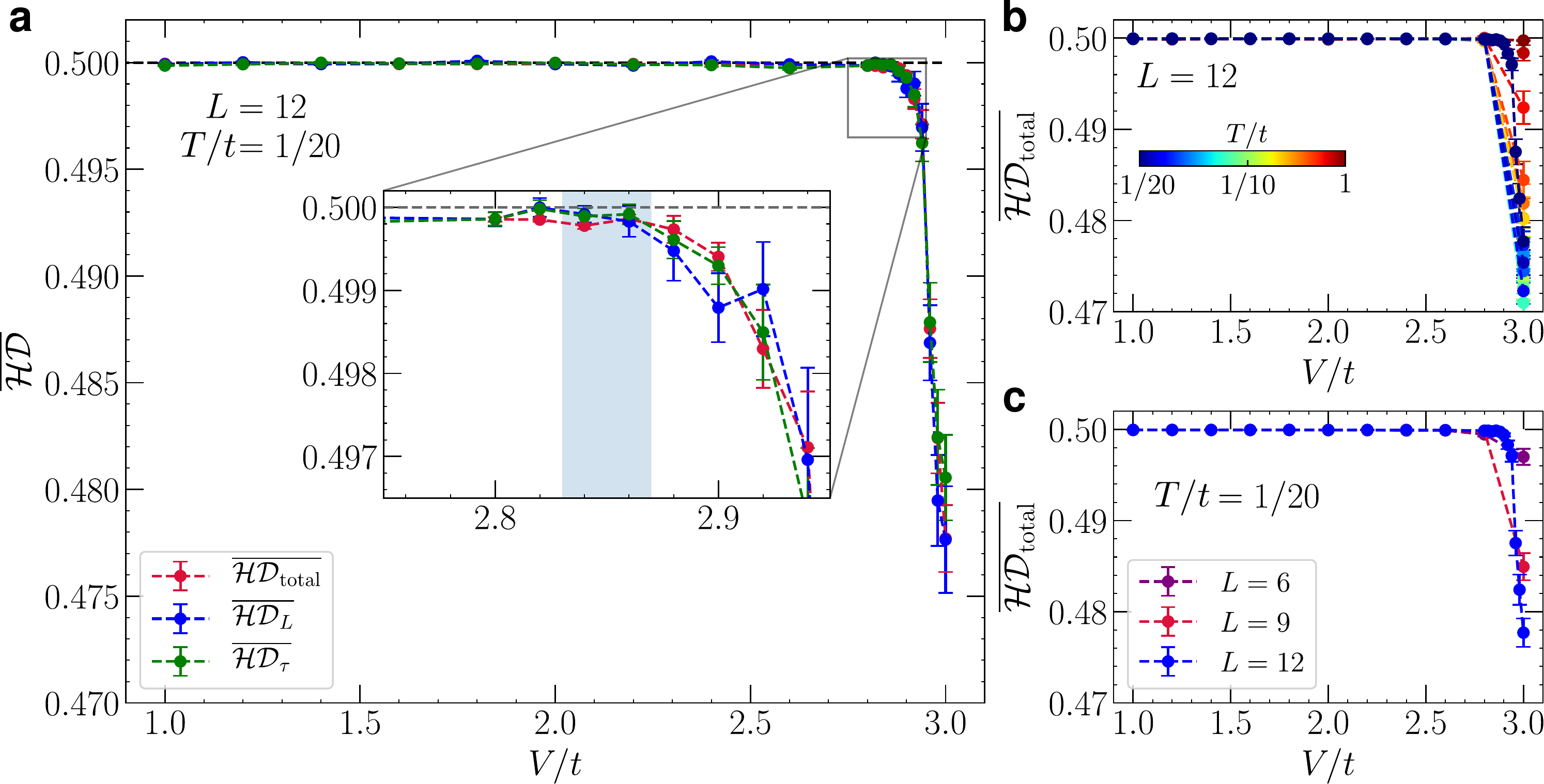}
\caption{\textbf{Hamming distance for the spinless triangular Hubbard model} \textbf{a,} The different averaged Hamming distances with increasing repulsive nearest-neighbor interactions $V$. The inset displays a zoom-in of the location where ${\cal HD}$ departs from 1/2, accompanied by a shaded region marking a confidence interval of the ordered phase onset for an $L=12$ lattice at $T/t = 1/20$. \textbf{b,} shows the total Hamming distance with decreasing temperature, while \textbf{c,} displays its finite-size effects at $T/t=1/20$. ED results give a transition at $V/t \simeq 2.7$ on an $L=6$ lattice~\cite{SI}.}
\label{fig:Fig4}
\end{figure}

\paragraph{The U(1) triangular Hubbard model.---} Building on those results, we investigate a model in which there is no known solution to circumvent the sign problem, i.e., the geometrically frustrated triangular lattice Hubbard model. Starting from its spinless formulation, we notice that in between commensurate densities, as 1/3 and 2/3 fillings, a \textit{pinball liquid} phase arises, in which CDW order coexists with metallic behavior~\cite{Hotta2006,Miyazaki2009}; this intermediate phase corresponds to a supersolid in the case of hardcore bosons~\cite{Wessel2005}. With the goal of exploring a well marked QCP, we study the case of one fermion for every three lattice sites.

Even though the specific location of the critical interaction strength that leads to a form of Wigner solid is currently not known, the average Hamming distance analysis when sweeping $V$ shows a well marked transition at $V_c/t = 2.85(2)$ (Fig.~\ref{fig:Fig4}\textbf{a}). ED results (see SI~\cite{SI}) in smaller lattices ($L=6$) can capture a continuous transition at $V/t \simeq 2.7$ instead, highlighting the importance of finite-size effects in determining the QCP location. Temperature and lattice size dependence in these results are displayed in Figs.~\ref{fig:Fig4}\textbf{b} and \ref{fig:Fig4}\textbf{c}, respectively. The former shows a subtle non-monotonic behavior of $\overline{\cal HD}$ within the ordered phase when $T$ is decreased, whose origin will be explored in detail for the spinful version in what follows.

\paragraph{The SU(2) triangular Hubbard model.---} The ingredient that allows both insulating and antiferromagnetic transitions to concomitantly occur on the SU(2) Hubbard model with growing interactions, i.e., that the cluster structure is bipartite, is no longer present in a triangular lattice. As a result, early studies within approaches that try to conquer the first exponential `wall' managed to demonstrate the existence of an intermediate non-magnetic insulating (NMI) phase, separating the metallic regime at small interactions and the magnetically ordered phase at large values of $U/t$~\cite{Sahebsara2008,Yoshioka2009,Yoshioka2009,Laubach2015}. In the latter, the low-energy effective model maps to the antiferromagnetic Heisenberg model, in which a $120^{\circ}$ Néel ordered phase has been shown to be stabilized to compose with the geometric frustration~\cite{Huse1988,White2007}.

\begin{figure}[t!]
\centering
\includegraphics[width=0.99\columnwidth]{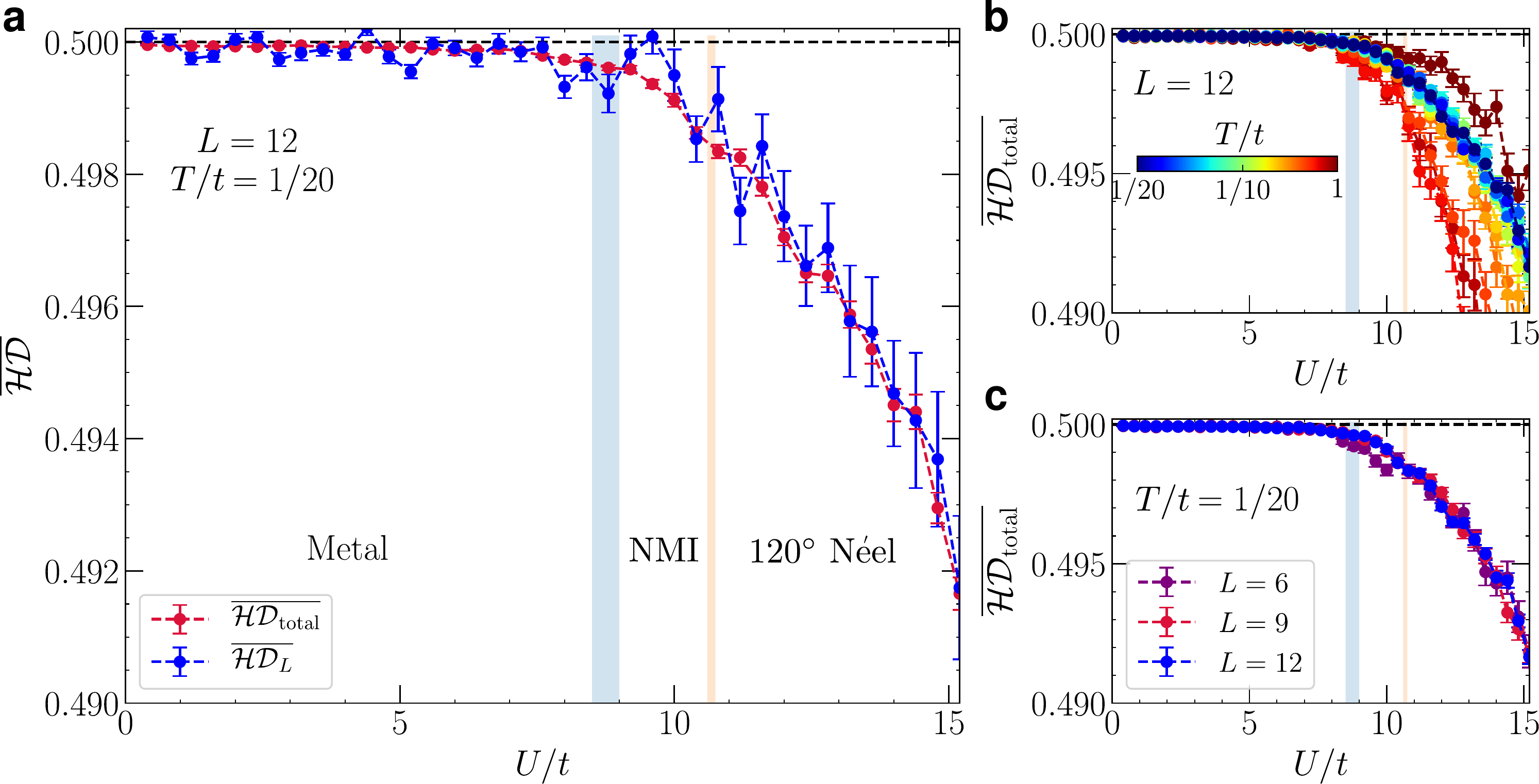}
\caption{\textbf{Hamming distance for the spinful triangular Hubbard model.} \textbf{a,}  Similar to Fig.~\ref{fig:Fig2}, but for the SU(2) Hubbard model on the triangular lattice. \textbf{a,} Total and $\tau=\beta/2$ average Hamming distances; \textbf{b,} Temperature dependence of the Hamming distance for a fixed lattice size ($N_s = 144$) and, \textbf{c,} the finite-size effects obtained at $T/t = 0.05$. Shaded vertical regions in all panels display the combined results from Refs.~\cite{Szasz2020,Chen2021} obtained for width-4 cylinders with finite or infinite lengths; they mark transitions at $U_{c1}/t = 8.5 - 9$ and $U_{c2}/t = 10.6-10.75$.}
\label{fig:Fig5}
\end{figure}

The nature of the intermediate NMI phase has been investigated within density matrix renormalization group (DMRG) methods~\cite{Shirakawa2017,Szasz2020,Wietek2021,Chen2021}, and recent results have pointed out to the possibility that this phase realizes a gapped chiral spin liquid~\cite{Szasz2020,Chen2021}, at least in some of the lattice structures amenable to computations. Importantly, molecular crystals of the $\kappa$-ET family are known to be close experimental realizations of such triangular lattice Hubbard models~\cite{Kanoda2011,Powell2011}, and in particular, compounds as $\kappa$-(ET)$_2$Cu$_2$(CN)$_3$ have a quasi-isotropic hopping structure in the lattice~\cite{Shimizu2003}, which does not exhibit any magnetic ordering down to 32mK, indicative that it may indeed host a spin liquid ground state.

Despite the strong motivation provided by these exciting results, progress has been significantly impeded because the corresponding model has a drastic sign problem, and QMC calculations have
large error bars stemming from the second exponential wall we described in the introduction. Nevertheless, as Fig.~\ref{fig:Fig5} shows, although extremely challenging to extract physical quantities, the onset of the ordered phase can be very successfully observed via the average Hamming distance. The deviation from the uncorrelated case ($\overline{\cal HD} = 1/2$) is seen to be reasonably well aligned to the most recent predictions on this model~\cite{Szasz2020,Chen2021}.

\begin{figure}[htp]
\centering
\includegraphics[width=0.9\columnwidth]{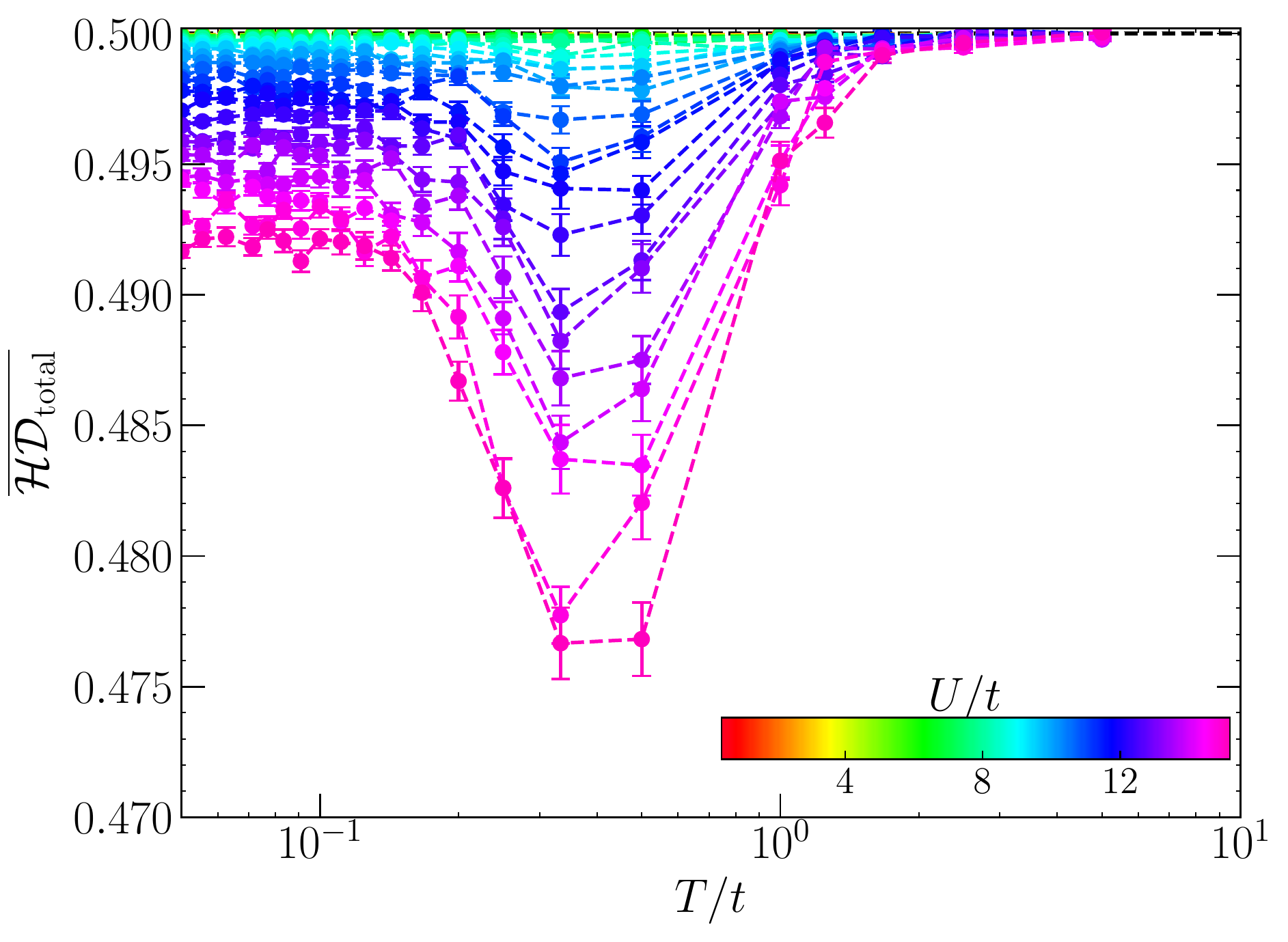}
\caption{\textbf{Order-by-disorder effects in the Hamming distance.} Hamming distance for the spinful triangular Hubbard model {\it{v.s.}} $T$ for various interaction strengths $U$ in an $L=12$ lattice. The decrease in the Hamming distance upon heating, that is, the increase of localization in the phase space, parallels similar effect observed for a physical quantity, the double occupancy, a proxy for electronic localization~\cite{Laubach2015,Wietek2021}, whose minima occur at compatible temperatures.}
\label{fig:Fig6}
\end{figure}

Although a clear-cut location of the QCP is likely only obtained in the $\Delta\tau\to0$ limit (see such analysis for the U(1) honeycomb Hubbard model in SI~\cite{SI}), there is manifest evidence that the Hamming distance does capture physically relevant information. An argument in this direction can be put forward by early observations that point out that the interplay of geometric frustration and interactions may lead to ground states with large thermal entropies $S$ when entering the ordered regime. In turn, the positive variation of $S$ with interactions, at fixed $T$, can be related to the decrease of the double occupancy $D = (1/N_s)\sum_{i} \langle \hat n_{i\uparrow} \hat n_{i\downarrow}\rangle$ with temperature (at fixed $U$) via a Maxwell relation~\cite{Li2014,Laubach2015,Wietek2021}
\begin{equation}
\frac{\partial S}{\partial U}\bigg\rvert_{T} = -\frac{\partial D}{\partial T}\bigg\rvert_{U}.
\end{equation}
As a consequence, the double occupancy at sufficiently small $T$'s decreases with temperature, an effect at odds of what one would expect from the connection of $D$ with localization (in the $U/t\to\infty$ limit, $D\to0$). This increase of electron localization upon heating, referred as order-by-disorder, can be similarly seen in Fig.~\ref{fig:Fig6}, by means of localization in phase space described by a reduced Hamming distance. A further confirmation of the unexpected $D \leftrightarrow {\cal HD}$ connection can be made by noticing that the minima of both quantities are seen at similar temperatures $T/t\approx 0.5$~\cite{Wietek2021}.

\paragraph{Discussion and outlook.---} A common thread in the study of quantum correlated matter is that if a given model is plagued by the sign problem when utilizing a QMC method, and the aim is to extract properties close to $T=0$, there is not much one can do, and the application of other numerical methods would consist of a better use of resources. This reasoning is based on the computationally expensive `wall' that prevents a statistically convergent estimation of physical quantities within reasonable time. What we have shown here is that other statistical metrics host information about the onset of ordered phases. This is clearly seen via the average distance in phase space spanned on the course of the importance sampling, but other complementing metrics might reveal even finer details~\cite{Tiago2021a,Tiago2021b}.

Models featuring deconfined quantum critical points (DQCPs)~\cite{Senthil2004}, that is, separating two symmetry-incompatible ordered phases~\cite{Li2017}, are likely challenging to investigate using our approach. Nonetheless, a possible approach is to compare the average Hamming distance between configurations using slightly different Hamiltonian parameters. A likely outcome is that at sufficiently low temperatures, typical configurations sampled in parallel are on average far apart if the parameters are chosen such that they belong to different ordered phases. By systematically reducing the parameter's deviation, the point at which the ${\cal HD}$ decreases potentially signals the DQCP location. We leave this line of inquiry for future studies.

Lastly, it remains to be seen if the highly non-local action for the resulting HS fields that arises after the fermionic integration may define a spin glass at sufficiently low temperatures, and its eventual connection to negative weight configurations. The Hamming distance, intimately related to the spin-glass order parameter~\cite{Parisi1983}, has been used to quantify the ultrametricity of the phase space in classical models~\cite{Katzgraber2009}, a characteristic feature of glassy behavior~\cite{MezardParisiVirasoro1986}.

\paragraph{\textbf{Methods}}
Numerical calculations employ the Blanckenbecler-Scalapino-Sugar (BSS) algorithm~\cite{Blankenbecler1981,Hirsch1985} for QMC calculations: The partition is written as a path integral, in which a sequence of Trotter decomposition, HS transformation, and integration of the resulting fermionic bilinear forms allows one to express it as a sum over auxiliary field configurations of determinants of fermionic matrices on a single-particle basis. The sign problem thus arise as the determinants are not guaranteed to be positive definite for arbitrary configurations of the fields. We apply the standard spin-decomposition in the HS transformation for the SU(2) models~\cite{Hirsch1983},
\begin{align}
e^{-\Delta \tau U (\hat n_{i\uparrow} - \frac{1}{2}) 
(\hat n_{i\downarrow} - \frac{1}{2})} 
= \frac{1}{2} e^{-U \Delta \tau/4} \sum_{s_i=\pm 1} e^{\lambda s_i (\hat n_{i\uparrow} - \hat n_{i\downarrow})}
\, , \nonumber
\end{align}
or its corresponding on the U(1) Hamiltonians~\cite{Buendia1986},
\begin{align}
e^{-\Delta \tau V (\hat n_i - \frac{1}{2}) (\hat n_j - \frac{1}{2})} 
= \frac{1}{2} e^{-V \Delta \tau/4} \sum_{s_{ij}=\pm 1} e^{\lambda s_{ij} (\hat n_i - \hat n_j)}, \nonumber
\end{align}
where ${\rm cosh} \, \lambda = e^{U \Delta \tau / 2}$ and ${\rm cosh} \, \lambda = e^{V \Delta \tau / 2}$, respectively.
Note that the auxiliary fields that decouple the interactions have a double index in the latter, and reside on the bonds connecting orbitals $i$ and $j$. The total number of bonds, and correspondingly number of auxiliary field configurations in a single imaginary-time slice is $N_b = 3L^2$ in either the honeycomb or triangular lattices. We do not make use of the Majorana representation~\cite{Li2015}, thus our simulations are affected by the sign problem in the U(1) honeycomb Hubbard model, which is irrelevant for our results, and highlight the predictive power of statistical properties of the importance sampling. Simulations are carried out employing typically thousands of QMC sweeps with independent Markov chains ranging from 20 to 48. In the SI~\cite{SI}, we further employ ED for the triangular Hubbard model in small clusters, featuring 36 (18) sites in its spinless (spinful) formulation.

{\bf Acknowledgements:} R.M.~acknowledges support from the National Natural Science Foundation of China (NSFC) Grants No. U1930402, 12050410263, 12111530010 and No. 11974039. 
R.T.S. was supported by the grant DE‐SC0014671 funded by the U.S. Department of Energy, Office of Science. 
Computations were performed on the Tianhe-2JK at the Beijing Computational Science Research Center. 
{\bf Author contributions:} R.M.~designed the project and performed ED calculations, T.~Y.~undertook the QMC simulations. All authors analyzed the data and interpreted the results. {\bf Competing interests:} Authors declare no competing interests. {\bf Data and materials availability:} All data needed to reproduce the conclusions in the paper are present in the paper or the Supplementary Materials. Data presented in the figures are deposited at \cite{Zenodo}.

\bibliography{references}

\clearpage


\begin{center}

{\large \bf Supplementary Information:
 \\Quantum critical points and phase space distances in quantum Monte Carlo simulations}\\

\vspace{0.3cm}

\end{center}

\vspace{0.6cm}

\beginsupplement

\begin{figure}[htp]
\centering
\includegraphics[width=0.5\textwidth]{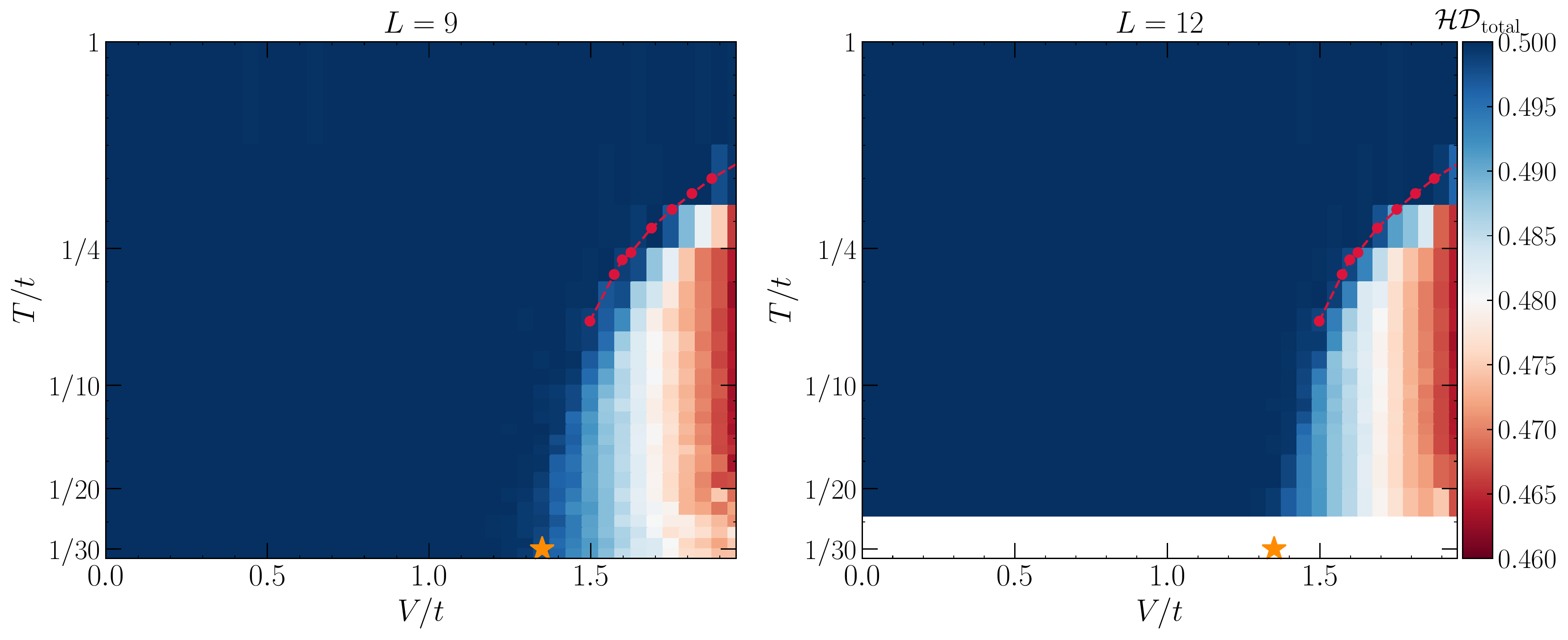}
\caption{\textbf{Thermal transitions via the Hamming distance.} Contour plot of the average Hamming distance in the temperature--interactions plane for the $t$-$V$ Hubbard model on the honeycomb lattice, with linear size $L=9$ (left) and $L=12$ (right). Markers are extracted from the results of Ref.~\cite{Hesselmann2016}, and star at the lowest temperature describes the known QCP location~\cite{LeiWang2014,Li2015}.
}
\label{fig:Fig_S1}
\end{figure}

\paragraph{Finite-temperature transitions.---} \label{sec:finiteT}
The U(1) Hubbard model displays a finite-temperature transition to an ordered phase, which has been classified on the honeycomb lattice by means of the continuous-time interaction expansion method~\cite{Hesselmann2016} or in a hybrid SSE/determinantal approach~\cite{Wang2016}, both of which result in sign-problem free simulations. Here we argue that the metric of phase space exploration we introduced in the main text, the average Hamming distance, similarly captures the thermal transition in the original BSS type algorithm for this model~\cite{Buendia1986}. Figure~\ref{fig:Fig_S1} displays the `phase diagram' of this quantity in the temperature--interactions plane, overlaying it with the results extracted from Ref.~\cite{Hesselmann2016}. Agreement between these results and the $(T,V)$ parameters at which $\overline{\cal HD}$ deviates from $1/2$ is reasonably good, presenting an even closer matching when increasing the system size.  

\begin{figure}[t!]
\centering
\includegraphics[width=0.99\columnwidth]{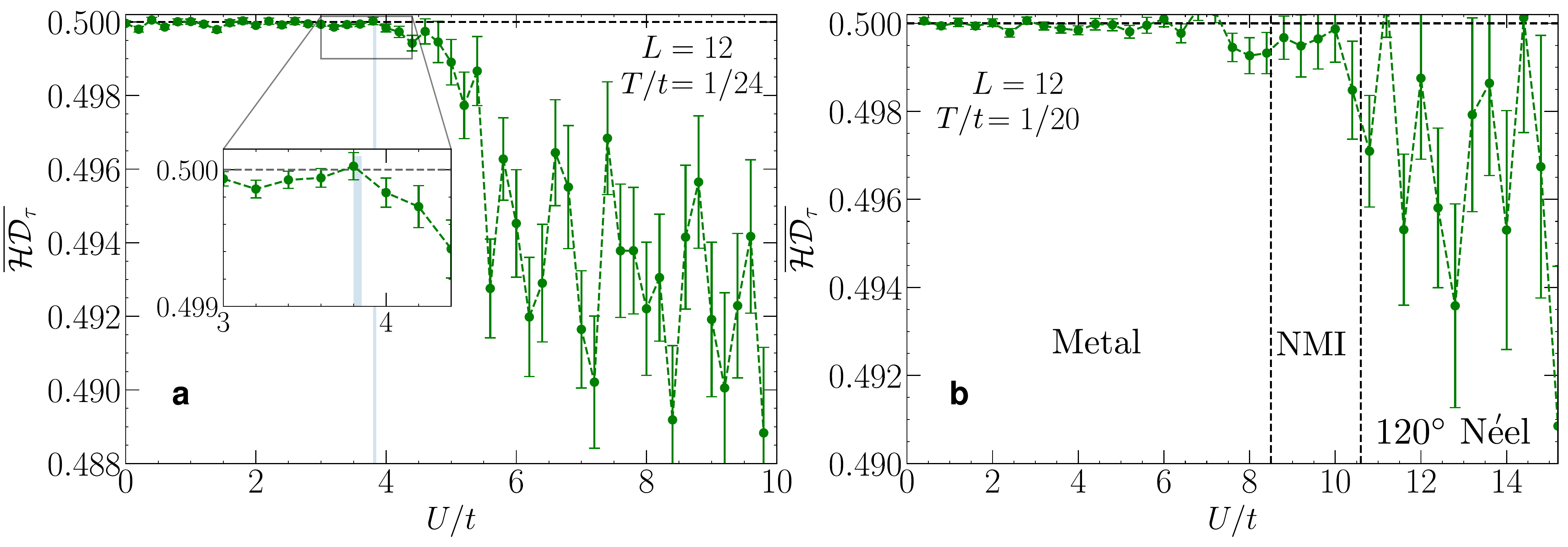}
\caption{\textbf{Fluctuations of the Hamming distance across imaginary-time.}
\textbf{a,} Average Hamming distance $\overline{{\cal HD}_{\tau}}$ in the spinful honeycomb Hubbard model while sweeping the interactions, and \textbf{b,} the same but for the spinful triangular Hubbard model. Although fluctuations are much larger than the other metrics we investigate, it qualitatively captures the onset of the ordered phase. In both cases the linear lattice size is $L=12$; the inverse temperature is set at $\beta t = 24$ and 20, respectively.}
\label{fig:HD_tau_spinful}
\end{figure}

\paragraph{Large statistical fluctuations in ${\cal HD}_\tau$.---} As described in the main text, statistical fluctuations in the average Hamming distance within a site (or unit cell) across different imaginary time-slices, ${\cal HD}_\tau$, are sensibly large, as seen in Fig.~\ref{fig:HD_tau_spinful} for the two SU(2) models we investigate. There are two reasons that can explain such behavior. The first is that the HS field is a massless bosonic field that mediates instantaneous interactions between the fermions traversing the real-space imaginary-time lattice; a consequence of being massless is that wild fluctuations occur in imaginary-time (even if the field were to be made continuous~\cite{Loh1992}). A direct contrast is the case of the Holstein model, where massive phonons play the role that mediate the electronic interactions. In this case, their mass controls the `velocity' of the phonon field in the action, taming the large oscillations in imaginary-time. In the absence of such kinetic energy term in the action for the Hubbard model, the field configurations in consecutive imaginary-time slices are not directly coupled. Second, spin orientation patterns for the fermions in approaching the atomic limit are more easily seen through equal-time correlations. Consequently, selecting one unit cell or site to monitor the auxiliary-field configurations, does not render a well defined string that uniquely captures such patterns owing to quantum fluctuations which inherently occur.

\paragraph{Approaching a continuous imaginary-time.---}
Apart from statistical uncertainties, which are always controllable by the sampling extent, certain QMC simulations (auxiliary-field versions, in particular) are subjected to a single approximation, that stems from the Trotter decomposition employed when splitting the exponential operators in the partition function. For the Hubbard model, for example, the simplest decomposition leads to an error $O[tU(\Delta\tau)^2]$. Nonetheless, this approximation can be made controllable by taking the limit $\Delta\tau \to 0$ (at the expense of increasing the number of imaginary-time slices at a fixed temperature) such that it becomes indistinguishable from the statistical fluctuations.

Here in the case of the statistics of the sampling, namely the average Hamming distance, there is another complication that was exposed in the main text when describing the connection between fermionic correlations and correlations between the auxiliary field components. This can be seen via the relation connecting bosonic field correlations and the fermionic ones~\cite{Hirsch1983,Hirsch1986}
\begin{equation}
    \langle [\hat n_{i\uparrow}(\tau) - \hat n_{i\downarrow}(\tau)][\hat n_{j\uparrow}(0) - \hat n_{j\downarrow}(0)]\rangle = \frac{1}{1 - e^{-\Delta\tau U}}\langle s_{i,\tau} s_{j,0}\rangle ,
\label{eq:Hirsch_eq}
\end{equation}
with the exception $i=j$, $\tau = 0$. As we argued there, for the proportionality constant to approach 1 in the atomic limit, it is necessary that the imaginary-time discretization to go to zero slower than that. As a result, a relevant analysis is to understand the effect of the imaginary-time discretization on the average Hamming distance. We do so for the U(1) honeycomb Hubbard model, where a crisp connection of the correlated sampling to the onset of the ordered phase was drawn. We notice that in this case of a spinless Hamiltonian, a similar relation as Eq.~\eqref{eq:Hirsch_eq} can be derived, involving the correlation of density operators $(\hat n_i - \hat n_j)$ and the corresponding decoupling field $s_{ij,\tau}$ at that bond.

\begin{figure}[t!]
\includegraphics[width=0.9\columnwidth]{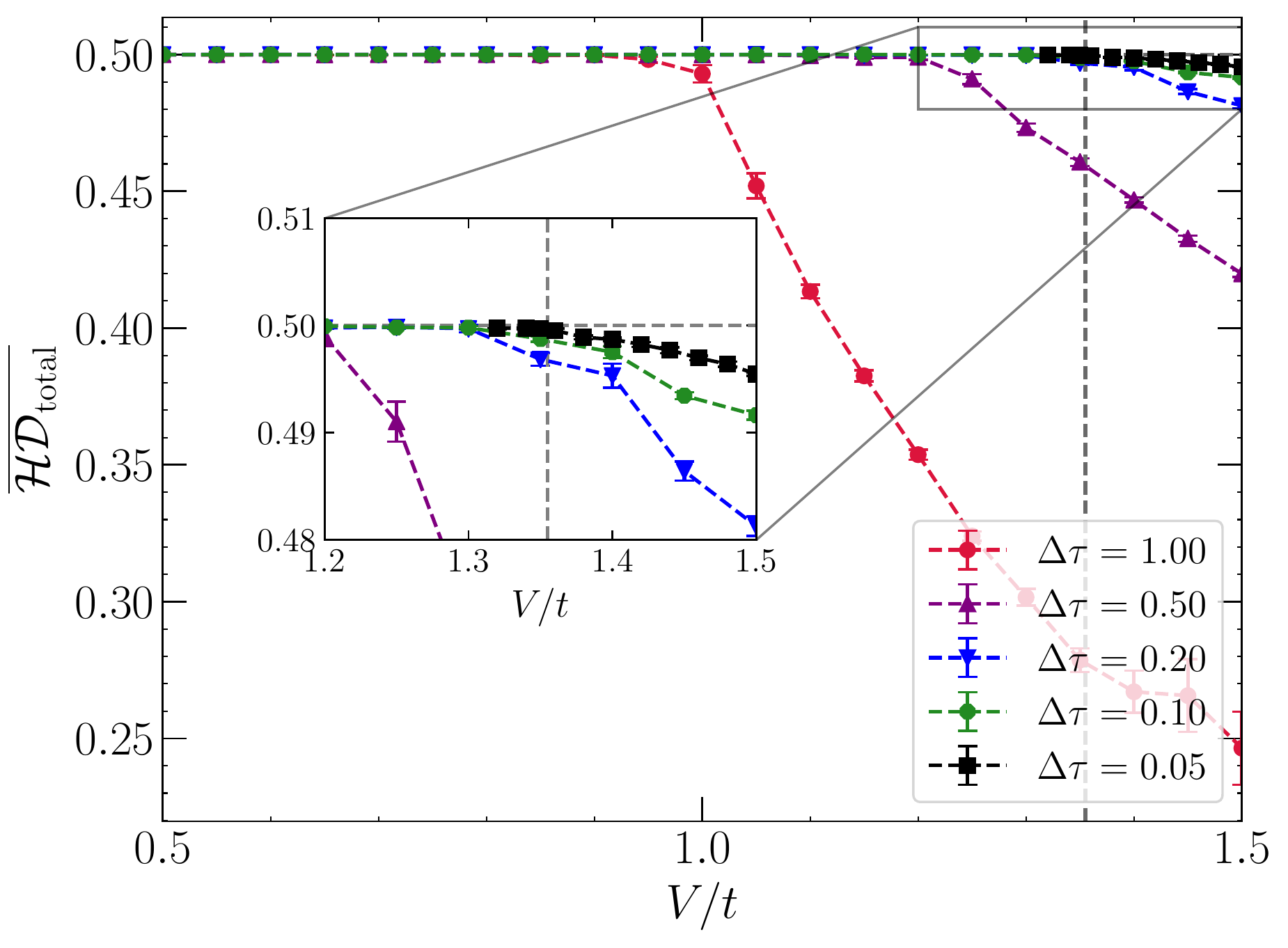}
\caption{\textbf{Hamming distance dependence on the imaginary-time discretization.} Total Hamming distance $\overline{{\cal HD}_{\rm total}}$ in the spinless honeycomb Hubbard model with a range of imaginary-time discretizations $
\Delta\tau$ as marked. Vertical dashed lines give the known critical transition $V_c = 1.355t$~\cite{Li2015, LeiWang2014} related to the onset of the ordered phase. The inset gives a zoom-in depicting the asymptotic approach to the transition. Here the linear lattice size is $L=9$, and the temperature is fixed at $T/t = 1/16$.
}
\label{fig:HD_dtau}
\end{figure}

For that end, Fig.~\ref{fig:HD_dtau} shows the average total Hamming distance with decreasing $\Delta\tau$, at a low fixed temperature $T/t = 1/16$ on an $L=9$ lattice. An asymptotic approach to the known critical point is obtained in the limit $\Delta\tau\to 0$, and values of $\Delta\tau=0.1$ are sufficiently close to describe the onset of the ordered phase.

\begin{figure}[t!]
\includegraphics[width=0.9\columnwidth]{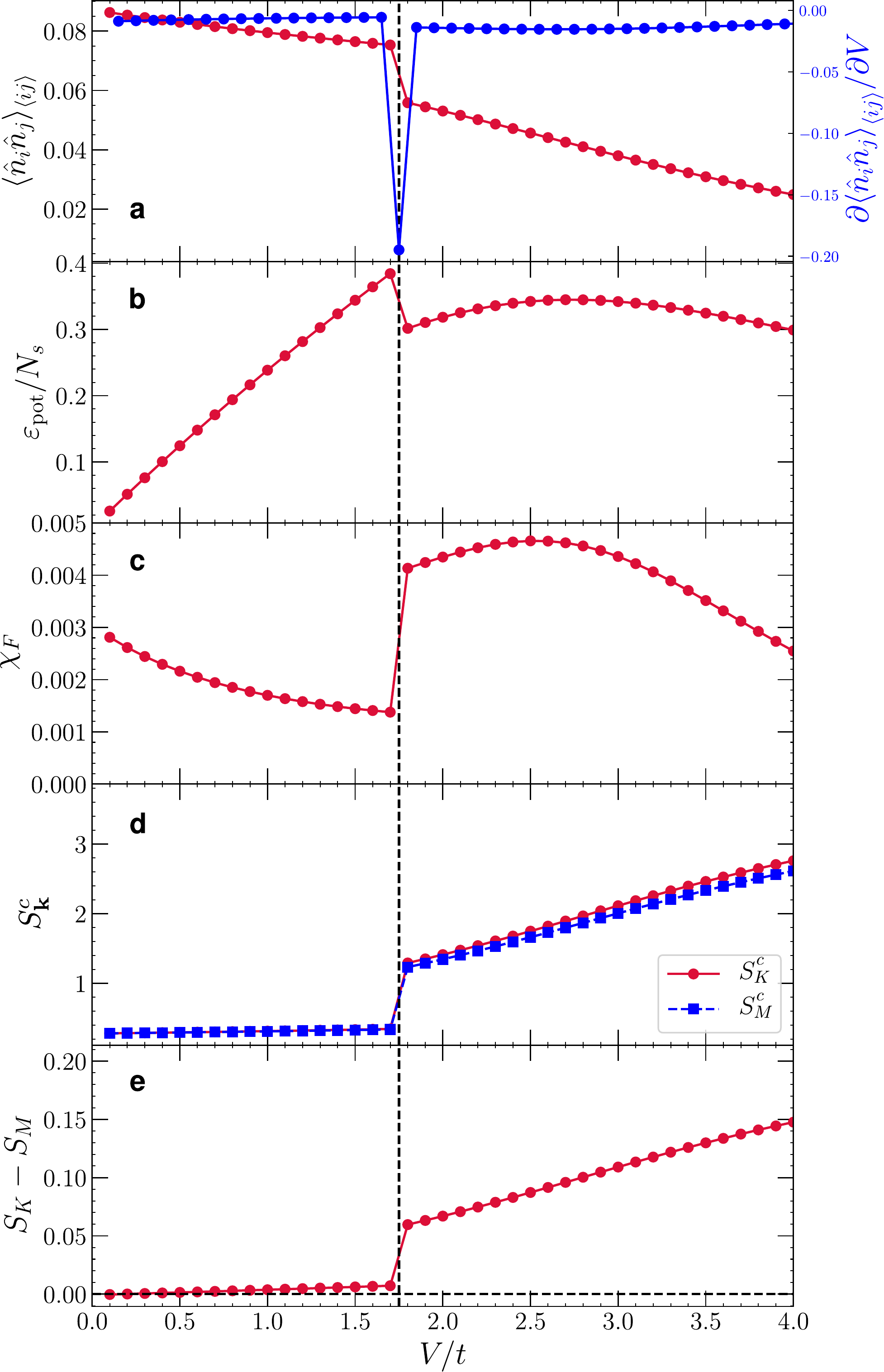}
\caption{\textbf{ED results for the spinless triangular Hubbard model on a $6\times 6$ lattice.} \textbf{a,} Average nearest-neighbor correlation function (left $y$-axis) and its corresponding derivative (right $y$-axis). \textbf{b,} The vertical dashed lines across all panels mark the location of the first order phase transition observed for this cluster size, where the momentum sector ground state resides changes with $V$.}
\label{fig:ED_spinless}
\end{figure}

\begin{figure}[t!]
\includegraphics[width=0.85\columnwidth]{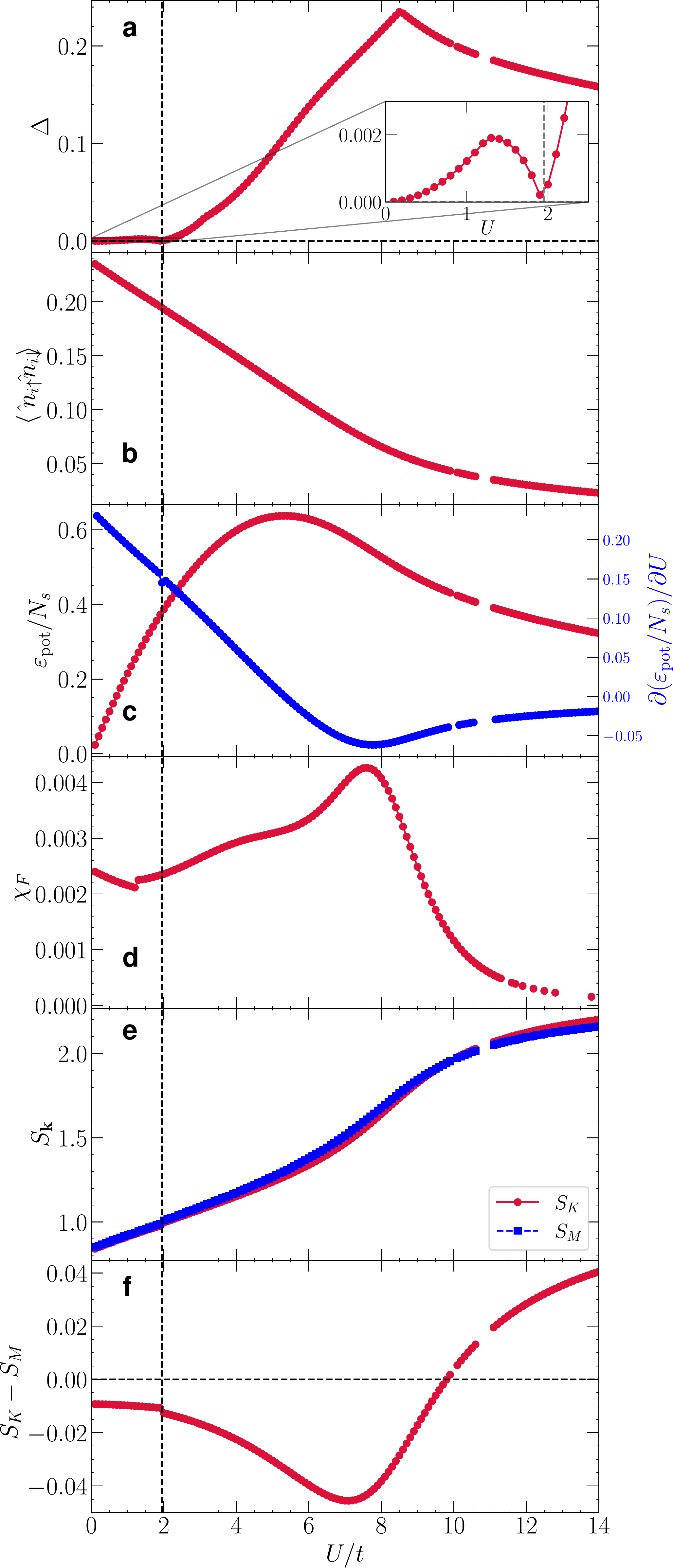}
\caption{\textbf{ED results for the spinful triangular Hubbard model on a $6\times 3$ cluster.} \textbf{a,} The many-body gap $\Delta = |E^{\rm GS}_{(0,0)} -E^{\rm GS}_{(0,4\pi/3)}|$ comparing the lowest energy states at different momentum sectors. The inset magnifies the results showing a first order phase transition occurring at $U/t \simeq 1.9$. \textbf{b,} The double occupancy at the ground state, and \textbf{c,} the potential energy vs. $U/t$. The inflection point in the latter likely gives the metal-insulating transition point ($U/t \simeq 7.6$) which coincides with the peak of the fidelity susceptibility shown in \textbf{d}. \textbf{e,} The spin structure factor at different high symmetry points, and \textbf{f,} the difference of the results for $K$ and $M$ points in the Brillouin zone. They point out to an enhanced stripy antiferromagnetic correlations which gives way to a Néel 120$^\circ$ order at $U/t \simeq 10$ (see text).}
\label{fig:ED_spinful}  
\end{figure}

\paragraph{Exact diagonalization results.---} Despite being a useful tool to infer the onset of ordered phases, the average Hamming distance does not aid in understanding the precise nature of those phases. The honeycomb lattice models we study are well understood, but the triangular lattice versions are slightly less explored. For that reason, we make use of exact diagonalization (ED) to corroborate some of these phase transitions and ensuing ordered phases. Constrained by the first exponential wall, we are limited to calculations on small cluster sizes, which we set at $6\times 6$ and $6 \times 3$ in the spinless and spinful Hamiltonians, respectively.

We start with the U(1) case, with filling $1/3$ (12 fermions in 36 orbitals), as shown in Fig.~\ref{fig:ED_spinless}. When sweeping the interactions $V$ we observe a first order phase transition at around $V/t\simeq 1.7$, where the ground state changes from the pseudomomentum $\vec q = (0, 2\pi/3)$ sector to $\vec q = (0, 0)$. This leads to discontinuities in several observables, including the average nearest-neighbor density correlations $\langle \hat n_i \hat n_j \rangle_{\langle i j \rangle}$ (Fig.~\ref{fig:ED_spinless}\textbf{a}). We believe that such a transition is cluster-dependent. More importantly, there is clear signal in physical quantities of an actual continuous phase transition at larger interactions, likely associated to the onset of the 1/3-CDW phase. A first indication is seen by means of the potential energy $\varepsilon_{\rm pot} = V\sum_{\langle i,j\rangle} \langle \hat n_i \hat n_j \rangle$, which displays a peak at around $V/t \simeq 2.7$ (Fig.~\ref{fig:ED_spinless}\textbf{b}); past this point one expects that quantum fluctuations are necessarily reduced in heading towards the atomic limit and $\varepsilon_{\rm pot}$  decreases. Similar information about the transition can be inferred by the fidelity susceptibility,
\begin{equation}
    \chi_F = \frac{2}{N_s} \frac{1 - |\langle \Psi_0(V)|\Psi_0(V+dV)}{dV^2},
\end{equation}
where $|\Psi_0(V)\rangle$ is the ground state with interaction magnitudes $V$, and $dV = 10^{-3}t$ is a small parameter deviation. Extensive peaks (with the system size) signify locations of QCP for continuous transitions, whereas discontinuities are seen within first-order ones. For our specific case of a 36-sites lattice, we observe a peak at around $V/t \simeq 2.5$ (Fig.~\ref{fig:ED_spinless}\textbf{c}).

A precise characterization of this phase can be obtained by the charge structure factor,
\begin{equation}
    S_{\bf k}^c = \frac{1}{N_s} \sum_{l,m}^{N_s} e^{i {\bf k}\cdot ({\bf r}_l - {\bf r}_m)} \langle \left(\hat n_l - 1/3 \right) \left(\hat n_m - 1/3 \right)\rangle.
\end{equation}
For ${\bf k} = (2\pi/\sqrt{3}, 2\pi/3) \equiv K$, a three-sublattice pattern occupancy is favored, whereas ${\bf k} = (2\pi/\sqrt{3},0) \equiv M$ describes a stripe density pattern. Figure~\ref{fig:ED_spinless}\textbf{d} shows the structure factor at these two points, where both grow with $V$, but the $K$-channel has always larger amplitude (Fig.~\ref{fig:ED_spinless}\textbf{d} and \textbf{e}). We thus conclude that there is a tendency of three-sublattice charge occupation, but a system-size scaling may clarify how these result on a finite-order parameter when approaching the thermodynamic limit.

We now describe the results of the spinful triangular Hubbard model at half-filling on a lattice with 18 sites (Fig.~\ref{fig:ED_spinful}). This lattice is similar to the YC3 cluster investigated within DMRG schemes~\cite{Szasz2020,Wietek2021}, albeit with longitudinal size $L_x = 6$. As for the U(1) case, there is also a change of the momentum sector associated to the ground state at a small cluster size, which leads to a first order phase transition at $U/t \simeq 1.9$. This is seen in Fig.~\ref{fig:ED_spinful}\textbf{a} which gives the gap between the $\vec q = (0,0)$ and $\vec{q} = (0, 4\pi/3)$ momentum sectors ground states, $\Delta \equiv |E^{\rm GS}_{(0,0)} -E^{\rm GS}_{(0,4\pi/3)}|$; initially at small interactions at $E^{\rm GS}_{(0,0)} > E^{\rm GS}_{(0,4\pi/3)}$, while $E^{\rm GS}_{(0,0)} < E^{\rm GS}_{(0,4\pi/3)}$ past $U/t \simeq 1.9$.

Nonetheless, at larger interactions a more smooth behavior is observed, where the fidelity susceptibility displays a clear peak at $U/t \simeq 7.6$ (Fig.~\ref{fig:ED_spinful}\textbf{d}), which coincides with the positions at which the potential energy $\varepsilon = U \sum_i \langle \hat n_{i\uparrow} \hat n_{i\downarrow}\rangle $ has an inflection point (Fig.~\ref{fig:ED_spinful}\textbf{c}). We thus believe this marks a regime where the metal-Mott insulating transition takes place. Recent estimations using width-4 cylinders~\cite{Szasz2020,Chen2021} put this transition in the range $U_{c1}/t = 8.5-9$. The nature of the intermediate region, dubbed a non-magnetic insulator, is under current debate, which has been converging towards a gapped chiral spin liquid phase. Yet, recent results using finite-temperature simulations~\cite{Wietek2021}, show that stripy antiferromagnetic spin correlations to be particularly pronounced there. For that reason, we compute the spin structure factor,
\begin{equation}
    S_{\bf k} = \frac{1}{N_s} \sum_{l,m}^{N_s} e^{i {\bf k}\cdot ({\bf r}_l - {\bf r}_m)} \langle \hat s_l^z \hat s_m^z\rangle,
\end{equation}
where $\hat s_l^z \equiv \hat n_{l\uparrow} - \hat n_{l\downarrow}$ (owing to the SU(2)-symmetric nature of the Hamiltonian, total spin structure can be obtained by multiplying the results by three). The results point out that although the spin structure factor at different momenta are all enhanced by the interactions (Fig.~\ref{fig:ED_spinful}\textbf{e}), at the $M$ point, which corresponds to the stripy antiferromagnetic spin correlations, a larger magnitude until $U/t \simeq 10$ indicates that such spin pattern is favored (Fig.~\ref{fig:ED_spinful}\textbf{f}). For values $U/t \gtrsim 10$, the $K$-spin structure factor dominates, and a 120$^\circ$ Néel spin ordered state likely takes place. Again, it remains to be seen how these results converge when a proper finite-size scaling is performed, which remains elusive with the limited system sizes amenable to exact calculations. 

\end{document}